\documentclass[twoside,american,english,sort&compress]{iopart}
\usepackage[T1]{fontenc}
\usepackage{geometry}
\usepackage{color}
\usepackage{babel}
\usepackage{array}
\usepackage{amsbsy}
\usepackage{amstext}
\usepackage{graphicx}
\usepackage[numbers]{natbib}
\usepackage[unicode=true,pdfusetitle,
 bookmarks=true,bookmarksnumbered=false,bookmarksopen=false,
 breaklinks=false,pdfborder={0 0 1},backref=false,colorlinks=true]
 {hyperref}
\hypersetup{
  citecolor=blue,linkcolor=blue,urlcolor=blue}

\makeatletter

\providecommand{\tabularnewline}{\\}

\usepackage{iopams}
\usepackage{setstack}

\newcommand{\eqref}[1]{(\ref{#1})}
\makeatother

\begin{document}
\title{Experiment data-driven modeling of tokamak discharge in EAST}
\author{Chenguang Wan$^{1,2}$, Jiangang Li$^{1,2}*$, Zhi Yu$^{1}$, and
Xiaojuan Liu$^{1}$}
\address{1. Institute of Plasma Physics Chinese Academy of Science, Hefei,
China}
\address{2. University of Science and Technology of China, Hefei, China}
\ead{\href{mailto:chenguang.wan@ipp.ac.cn}{chenguang.wan@ipp.ac.cn}}
\begin{abstract}
A model for tokamak discharge through deep learning has been done
on a superconducting long-pulse tokamak (EAST). This model can use
the control signals (i.e. Neutral Beam Injection (NBI), Ion Cyclotron
Resonance Heating (ICRH), etc) to model normal discharge without the
need for doing real experiments. By using the data-driven methodology,
we exploit the temporal sequence of control signals for a large set
of EAST discharges to develop a deep learning model for modeling discharge
diagnostic signals, such as electron density $n_{e}$, store energy
$W_{mhd}$ and loop voltage $V_{loop}$. Comparing the similar methodology,
we use Machine Learning techniques to develop the data-driven model
for discharge modeling rather than disruption prediction. Up to 95\%
similarity was achieved for $W_{mhd}$. The first try showed promising
results for modeling of tokamak discharge by using the data-driven
methodology. The data-driven methodology provides an alternative to
physical-driven modeling for tokamak discharge modeling.
\end{abstract}
\noindent{\it Keywords\/}: {tokamak, discharge modeling, machine learning}
\submitto{\NF }
\maketitle

\section{Introduction}

A reliable model of tokamak is critical to magnetic confinement fusion
experimental research. It is used to check the feasibility of pulse,
interpret experimental data, validate the theoretical model, and develop
control technology.

The conventional physical-driven modeling tools come from empirical
models or derivations based on first principles, the so-called \textquotedbl Integrated
Modeling\textquotedbl{} \citep{Falchetto2014}. \textquotedbl Integrated
Modeling\textquotedbl{} is a suite of module codes that address the
different physical processes in the tokamak, i.e. core transport,
equilibrium, stability, boundary physics, heating, fueling, and current
drive. Typical workflows are ETS \citep{Falchetto2014}, PTRANSP \citep{Budny2008},
TSC \citep{Kessel2006}, CRONOS \citep{Artaud2010}, JINTRAC \citep{Romanelli2014},
METIS \citep{Artaud2018}, ASTRA \citep{Pereverzev1991}, TOPICS \citep{Hayashi2010},
etc. The reliability of the first-principles model depends on the
completeness of the physical processes involved. In the past few decades,
sophisticated physical modules have been developed and integrated
into these codes for more realistic modeling results. Typical workflow
for a full discharge modeling on tokamak is using sophisticated modules
that integrate many physical processes \citep{Meneghini2015,Falchetto2014}.
Due to the nonlinear, multi-scale, multi-physics characteristics of
tokamak, high-fidelity simulation of the whole tokamak discharge process
is still a great scientific challenge \citep{Bonoli2015}.

Increasingly, researchers are turning to data-driven approaches. The
history can be traced back to the use of machine learning for interrupt
prediction since the 1990s, i.e. ADITYA \citep{sengupta2000forecasting,Sengupta2001},
Alcator C-Mod \citep{Rea2018,Tinguely2019,Montes2019}, EAST \citep{Montes2019,Zhu2020,Guo2020},
DIII-D \citep{Rea2018,Rea2018a,Montes2019,Wroblewski1997,Rea_2019,DeVries2011,Vega2013,Cannas2014,Ratta2014,Murari2018,Pau2018,Churchill2019,kates-harbeck2019be,Zhu2020}
JET \citep{Cannas2003,Vega2013,Windsor2005,Cannas2004,cannas2007support,Murari2008,Murari2009,Ratt2010,Ferreira2020,kates-harbeck2019be,Zhu2020},
ASDEX-Upgrade\citep{Cannas2010,pautasso2002line,Windsor2005,Aledda2015},
JT-60U \citep{Yoshino2003,Yoshino2005}, HL-2A \citep{yang2020,yang2020modeling},
NSTX \citep{Gerhardt2013} and J-TEXT \citep{Zheng2018,Wang2016}.
Neural-network-based models are also used to accelerate theory-based
modeling \citep{Honda2019,Meneghini2017,Meneghini2020}. In these
works \citep{Honda2019,Meneghini2017,Meneghini2020}, one neural network
was trained with a database of modeling and successfully reproduced
approximate results with several orders of magnitude speedup. There
are also many deep learning architectures have been created and successfully
applied in sequence learning problems \citep{Churchill2019,Graves2013,Lipton2015,IsmailFawaz2019,Ferreira2020}
in areas of time-series analysis or natural language processing. At
present, most machine learning work in fusion community estimates
the plasma state at each moment in time, usually identified as either
non-disruptive or disruptive. However, compared with traditional physical
modeling methods, it is far from enough to just predict whether the
disruption will occur. We need to understand the evolution of the
plasma state of tokamak and its response to external control during
the discharge process.

Physics-driven approaches reconstruct physical high-dimensional reality
from the bottom-up and then reduce them to the low-dimensional model.
Alternatively, data-driven approaches discover the relationships between
low-dimensional quantities from a large amount of data and then construct
approximate models of the nonlinear dynamical system. When focusing
only on the evolution of low-dimensional macroscopic features of complex
dynamic systems, data-driven approaches can build models more efficiently.
In practical applications, control signals and diagnostic signals
of tokamak usually appear as temporal sequence of low-dimensional
data, most of which are zero-dimensional or one-dimensional profiles
and rarely two-dimensional distribution. If we consider tokamak as
a black box, these signals can be considered as inputs and outputs
of a dynamic system. Discharge modeling is to model the connection
between the input and output signal. This can be understood as the
conversion from one kind of time series data to another.

In the present work, a neural network model is trained with the temporal
sequence of control signals and diagnostic signals for a large dataset
of EAST discharges \citep{Wan2015,Wan2019,ISI:000294731600008}. It
can reproduce the response of the main diagnostic signals to control
signals during the whole discharge process and predict their time
evolution curves, such as electron density $n_{e}$, store energy
$W_{mhd}$ and loop voltage $V_{loop}$.

The rest of this paper consists of five parts. Section \ref{sec:Dataset}
provides descriptions of data preprocessing and data selection criteria.
Section \ref{sec:Machine} shows the model details of this work. The
detailed model training process can be found in section \ref{sec:Training}.
Then an in-depth analysis of the model results is put forward in section
\ref{sec:Validation}. Finally, a brief conclusion is made in section
\ref{sec:Conclusion}.

\section{Dataset \label{sec:Dataset}}

EAST’s data system stores more than 3000 channels of raw acquisition
signals and thousands of processed physical analysis data \citep{wang2018studyof},
which record the entire process of tokamak discharge. These data can
be divided into three categories: configuration parameters, control
signals, and diagnostic signals. The configuration parameters describe
constants related to device construction, such as the shape of the
vacuum chamber, the position of the poloidal magnetic field (PF) coils,
etc. The control signals are the external constraints actively applied
to the magnetic field coil and auxiliary heating systems, such as
the current of PF coils, or the power of Lower Hybrid Wave (LHW),
etc. The diagnostic signals are the physics information passively
measured from the plasma, such as electron density $n_{e}$, or loop
voltage $V_{loop}$, etc. The configuration parameters will not change
during the experiment campaign, so there is no need to consider them
unless a cross-device model is built. The discharge modeling is essentially
a process of mapping control (input) signals to diagnostic (output)
signals.

In the present work, three signals that can represent the key characteristics
of discharge are selected as outputs, which are plasma stored energy
$W_{mhd}$, electron density $n_{e}$ and loop voltage $V_{loop}$.
The input signal should include all signals that may affect the output.
In this paper, ten types signals are selected as inputs, such as plasma
current $I_{p}$, central toroidal magnetic field $B_{t0}$, current
of PF coils, and power of LHW, etc. In principle, these signals can
be designed in the experimental proposal stage. Detailed information
about input and output signals are listed in table \ref{tab:Model-input-and}.
Some of these signals are processed signals with a clear physical
meaning, and others are unprocessed raw acquisition signals. As long
as the input signal contains information to determine the output,
whether it is a processed physical signal will not affect the modeling
result.

\begin{table}
{\small{}\caption{The list of signals. \label{tab:Model-input-and}}
}{\small\par}

{\small{}}%
\begin{tabular}{lll>{\raggedright}p{0.1\paperwidth}}
\hline 
{\small{}Signals} & {\small{}Physics meanings} & {\small{}Unit} & {\small{}Number of Channels}\tabularnewline
\hline 
\multicolumn{3}{l}{{\small{}Output Signals}} & {\small{}3}\tabularnewline
\hline 
{\small{}$n_{e}$} & {\small{}Electron density} & {\small{}$10^{19}m^{-3}$} & {\small{}1}\tabularnewline
{\small{}$V_{loop}$} & {\small{}Loop voltage} & {\small{}$V$} & {\small{}1}\tabularnewline
{\small{}$W_{mhd}$} & {\small{}Plasma stored energy} & {\small{}$J$} & {\small{}1}\tabularnewline
\hline 
\multicolumn{3}{l}{{\small{}Input Signals}} & {\small{}65}\tabularnewline
\hline 
{\small{}$I_{p}$} & {\small{}Plasma current} & {\small{}$A$} & {\small{}2}\tabularnewline
{\small{}PF} & {\small{}Current of Poloidal field (PF) coils} & {\small{}$A$} & {\small{}14}\tabularnewline
{\small{}$B_{t0}$} & {\small{}Toroidal magenetic field} & {\small{}$T$} & {\small{}1}\tabularnewline
{\small{}LHW} & {\small{}Power of Lower Hybrid Wave Current Drive and Heating System} & {\small{}$kW$} & {\small{}4}\tabularnewline
{\small{}NBI} & {\small{}Neutral Beam Injection System} & {\small{}Raw signal} & {\small{}8}\tabularnewline
{\small{}ICRH} & {\small{}Ion Cyclotron Resonance Heating System} & {\small{}Raw signal} & {\small{}16}\tabularnewline
{\small{}ECRH/ECCD} & {\small{}Electron Cyclotron Resonance Heating/Current Drive System} & {\small{}Raw signal} & {\small{}4}\tabularnewline
{\small{}GPS} & {\small{}Gas Puffing System} & {\small{}Raw signal} & {\small{}12}\tabularnewline
{\small{}SMBI} & {\small{}Supersonic Molecular Beam Injection} & {\small{}Raw signal} & {\small{}3}\tabularnewline
{\small{}PIS} & {\small{}Pellet Injection System} & {\small{}Raw signal} & {\small{}1}\tabularnewline
\hline 
\end{tabular}{\small\par}
\end{table}

Tokamak discharge is a complex nonlinear process, and there is no
simple way to determine the connection between the control signals
and the diagnostic signals. Therefore, the input data set covers most
of the control signals that can be stably obtained, and the redundant
signals are not identified and excluded. Determining the clear dependence
between control signals and diagnostic signals is one of the main
tasks of data-driven modeling, and it is also a direction worth exploring
in the future. This work focuses on verifying the feasibility of data-driven
modeling, and will not discuss this issue in depth.

In practical applications, there are significant differences in the
sampling rate of raw signals $R_{raw}^{i}$, where $i$ is the index
of signal. The input and output signal data sets need to be resampled
at a common sampling rate $R_{c}$ to ensure that the data points
of different signals are aligned at the same time. If $R_{raw}^{i}<R_{c}$,
the raw signal data needs to be interpolated to complement the time
series. If $R_{raw}^{i}>R_{c}$, the raw signal data needs to be smoothed
to eliminate high-frequency fluctuations.

The resampling rate $R_{c}$ depends on the time resolution of the
output signal, which refers to the time resolution of the physical
process of interest rather than the sampling rate of the raw experiment
signal. The accuracy of the reproduction of the physical process determines
the quality of the modeling. The size of the data set determines the
length of the model training time and the efficiency of modeling.
A lower resampling rate means lower time resolution and poor modeling
quality. However, a higher resampling rate means greater computing
resource requirements and lower modeling efficiency.

\begin{figure}
\includegraphics[width=0.5\paperwidth]{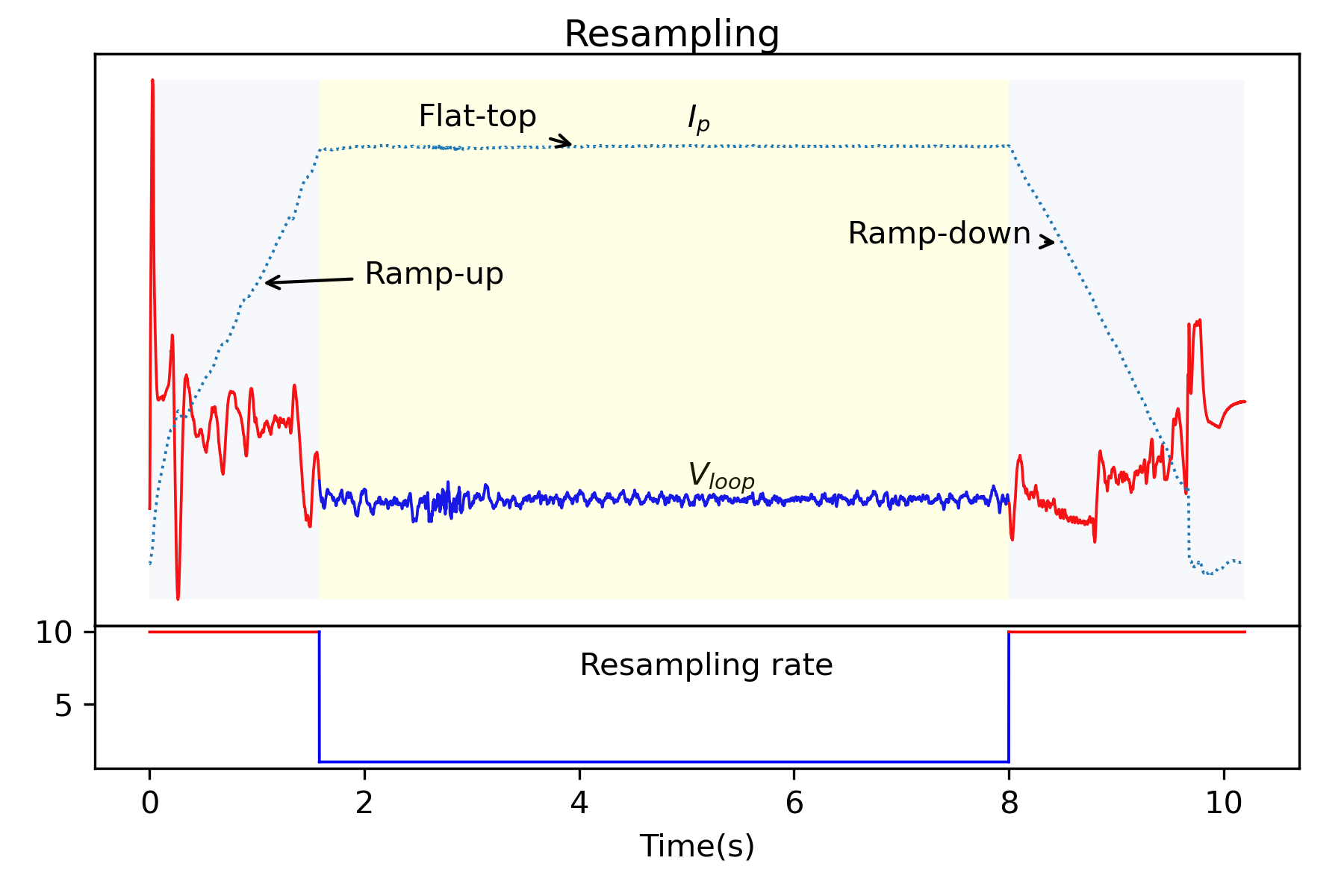}

\caption{Schematic diagram of adaptive resampling. The higher resampling rate
is used for segments that are of great interest to physicists or vary
greatly and frequently. \label{fig:Schematic}}
\end{figure}

The normal discharge waveform of the tokamak can be divided into three
phases, ramp-up, flat-top, and ramp-down (see figure \ref{fig:Schematic}).
Most signals climb slowly during the ramp-up phase, remain stable
during the flat-top phase, and slowly decrease during the ramp-down
phase. The time scale of the ramp-up and ramp-down phases are similar,
and the flat-top phase is much longer than the former two. The signals
show different time characteristics in these three phases. The signal
waveforms of $n_{e}$ and $W_{mhd}$ remain smooth in all three phases,
which can be accurately reproduced with a uniform resampling rate
$R_{c}^{n_{e}}=1kHz$. However, the waveform of $V_{loop}$ varies
greatly and frequently in the ramp-up and ramp-down phases, see Figure
\ref{fig:Schematic}. In order to ensure the quality of modeling,
the resampling rate of $V_{loop}$ is increased in these two phases
\[
R_{c}^{V_{loop}}=\left\{ \begin{array}{ll}
1kHz, & \text{flat-top},\\
10kHz, & \text{ramp-up or ramp-down }
\end{array}\right.,
\]
which is an adaptive piece-wise function. The purpose of using a non-uniform
adaptive resampling function is to balance the quality and efficiency
of modeling.

Three signal channels are selected as outputs, and 65 channels are
selected as inputs, and then resample according to the characteristics
of the output signal to align the time points. In the next step, machine
learning will be performed on these data.

\section{Machine learning architecture \label{sec:Machine}}

The data of tokamak diagnostic system are all temporal sequences,
and different signal data have different characteristics. According
to the temporal characteristics of the data of tokamak diagnostic
system, the sequence to sequence (seq2seq) model \citep{Sutskever2014}
was chosen as the machine learning model for tokamak discharge modeling.
Two methods of uniform resampling and adaptive resampling are adopted
for different data characteristic.

Natural language processing (NLP) is more similar to the discharge
modeling process than classification. NLP converts a natural language
sequence into another natural language sequence. It is an important
branch of machine learning, and the main algorithms are recurrent
neural network (RNN), gated recurrent unit (GRU) and long-term short-term
memory (LSTM), etc.

The encoder-decoder architecture \citep{Sutskever2014} is a useful
architecture in NLP, the architecture is partitioned into two parts,
the encoder and the decoder. The encoder’s role is to encode the inputs
into the state, which often contains several tensors. Then the state
is passed into the decoder to generate the outputs. In this paper,
LSTM was chosen as a fundamental component of the model. And stacked
LSTM as an encoder-decoder machine learning model.

\begin{figure}
\includegraphics[width=0.5\columnwidth]{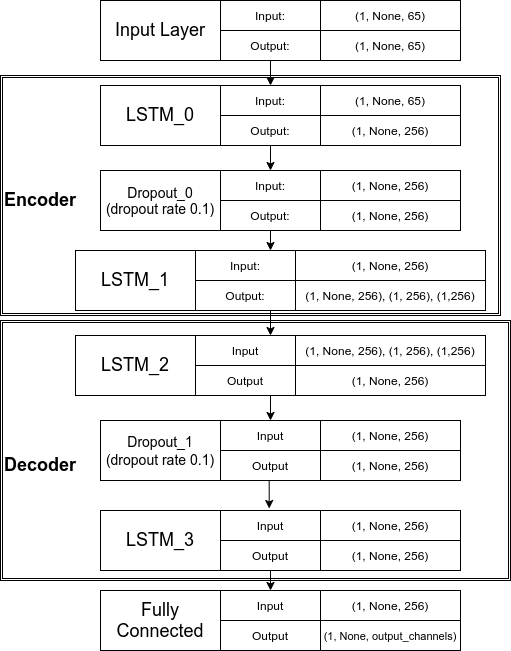}

\caption{Architecture of our model. Where \textquotedblleft None\textquotedblright{}
represents different lengths of sequence because of different discharge
shot duration time. \textquotedblleft output\_channels\textquotedblright{}
is the number of output sequences. \label{fig:Architecture}}
\end{figure}

As shown in figure \ref{fig:Architecture}, the machine learning model
architecture used in this work is based on the sequence to sequence
model (seq2seq) \citep{Sutskever2014}.   The first two LSTM layers
(LSTM\_0 and LSTM\_1 in figure \ref{fig:Architecture}) and Dropout\_0
can be considered as the encoder, and the last two LSTM layers (LSTM\_2
and LSTM\_3 in figure \ref{fig:Architecture}) and Dropout\_1 can
be regarded as the decoder. In this work, the encoder is to learn
the high-level representation (\emph{cannot be displayed directly})
of input signals (tab \ref{tab:Model-input-and} input signals). The
last hidden state of the encoder is used to \emph{initialize} the
hidden state of the decoder. The decoder plays the role of decoding
the information of the encoder. Encoder-decoder is built as an end-to-end
model, it can learn sequence information directly without manually
extracting features.

In terms of components, the main component of our architecture is
long short-term memory (LSTM) \citep{Hochreiter1997}, because the
LSTM can use trainable parameters to balance long-term and short-term
dependencies. This feature is suitable for tokamak sequence data,
tokamak discharge response is always strongly related to short-term
input changes but it is also affected by long-term input changes (e.g.
$W_{mhd}$ hardly changes fast, this property can be regarded as a
short-term dependence. However, the impact of other factors on energy
storage is cumulative. This property can be seen as long-term dependence.).
The dropout layer is a common trick to prevent over-fitting. The final
component is the fully connected layer to match the high-dimensional
decoder output with the real target dimension.

When considering the specific mathematical principles of the model,
the encoder hidden states $h_{t}$ are computed using this formula:

\begin{equation}
h_{t}=f(W^{(h_{lstm0}\delta_{dropout0}h_{lstm1})}h_{t-1}+W^{(hx)}x_{t}),\label{eq:1}
\end{equation}

where $h_{lstm0}$, $h_{lstm1}$ are the hidden state of LSTM \_0
and LSTM\_1 in figure \ref{fig:Architecture}, $\delta_{dropout0}$
is the dropout rate in Dropout\_0 in fig \ref{fig:Architecture},
$W$ and $W^{hx}$ are the appropriate weights to the previously hidden
state $h_{t-1}$ and the input vector $x_{t}$. $\delta_{dropout0}\sim Bernoulli(p)$.
This means $\delta_{dropout0}$ is equal to 1 with probability $p$
and 0 otherwise, we let $p=0.9$ for all experiment. Dropout\_0 means
that not all hidden states of LSTM\_0 can be transferred to LSTM\_1.

The encoder vector is the final hidden state produced from the encoder
part of the model. It is calculated using the formula above. This
vector aims to encapsulate the information for all input elements
to help the decoder make accurate predictions. It \emph{acts as the
initial hidden state} of the decoder part of the model.

In decoder a stack of two LSTM units where each predicts an output
$y_{t}$ at a time step $t$. Each LSTM unit accepts a hidden state
from the previous unit and produces and output as well as its hidden
state. In the modeling of tokamak discharge, The output sequence is
a collection of all time steps from the $y$. Any hidden state $h_{t}$
is computed using the formula:

\begin{equation}
h_{t}=f(W^{(h_{lstm2}\delta_{dropout1}h_{lstm3})}h_{t-1}),\label{eq:2}
\end{equation}

where $h_{lstm2,}h_{lstm3}$ are the hidden state of LSTM\_2 and LSTM\_3
in figure \ref{fig:Architecture}. $\delta_{dropout1}$ is the dropout
rate in Dropout\_1 in fig \ref{fig:Architecture}, $\delta_{dropout1}\sim Bernoulli(0.9)$.
$W$ is the appropriate weights to the previously hidden state $h_{t-1}$.
The process of data flowing in the decoder is similar to the encoder,
but the initial states of $h_{lstm2}$ is equal to the last states
of $h_{lstm1}$. As formula above, we are just using the previous
hidden state to compute the next one. The output $y_{t}$ at time
step t is computed using the formula:

\begin{equation}
y_{t}=activation(W^{D}h_{t}).\label{eq:3}
\end{equation}

The model calculates the outputs using the hidden state at the current
time step together with the respective weight $W^{D}$ of the fully
connected layer as shown in figure \ref{fig:Architecture} . The fully
connected layer is used to determine the final outputs. The activation
function is a linear function.

\section{Training \label{sec:Training}}

\begin{figure}
\includegraphics[width=0.5\textwidth]{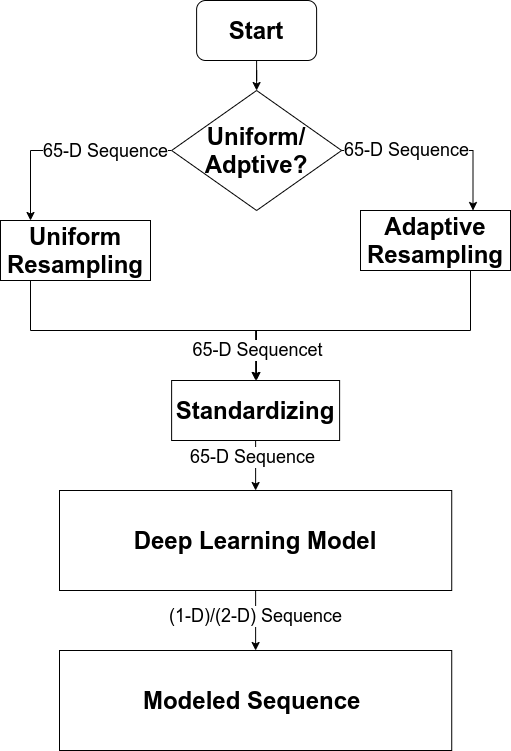}

\caption{Workflow of training. The resampling method is determined according
to the time characteristics of the output signal.\label{fig:workflow}}
\end{figure}

The form of the resampling function is determined according to the
time characteristics of the output signal waveform. The output signals
with the same sampling function are grouped together. The input signals
are resampled using the same sampling function as the output signal
set to ensure that all data points follow the same time axis. In this
section, model training and data processing will be introduced in
detail. The training of the model (see figure \ref{fig:workflow})
can be divided into five steps as follows:
\begin{enumerate}
\item Obtaining the data of 68 channels (include input and output signals
as shown in table \ref{tab:Model-input-and}) of the selected signals
from the EAST source database.
\item Using different resampling methods base on the time characteristics
of the output signal that would be modeled.
\item Standardizing the data with z-scores.
\item Data fed into the deep learning model for training.
\item Using the loss between model output and real experimental output as
backing propagation metric and then update parameters for the training
model.
\end{enumerate}
The dataset is selected from EAST campaign 2016-2018, discharge shot
number in the range \#70000-80000 \citep{Wan2015,Wan2013,ISI:000294731600008}.
A total 3476 normal shots are selected and divide the data into a
training set, a validation set, and a test set (training set: validation
set: test set = 6: 2: 2). The normal shot means that no disruption
occurred during this discharge, the flat top lasts more than two seconds,
and the key signals (i.e., model output signals, magnetic field, and
plasma current $I_{p}$) are complete. If there is no certain magnetic
field configuration, plasma is impossible to be constrained and without
completed $I_{p}$ or model output signals data there is no meaning
for tokamak discharge experiment or this model. The sampling of the
signal starts from $t=0s$ and continues to the end of the discharge
( typical EAST normal discharge time is five to eight seconds).

The shuffling method is a common improved generalization technique
\citep{KawaguchiLesliePackKaelbling} used in an entire data set.
However, in this work, the method is not used in the entire data set.
In order to prevent data leakage caused by multiple adjacent discharge
experimental shots with similar parameters, the entire data set is
divided into training set, validation set and test set according to
the experimental shot order. The phenomenon of adjacent discharge
experimental shots with similar parameters is very common in tokamak
discharge experiments. For generalization reasons, it is also necessary
to shuffle shot order in the training set. For example, there are
ten normal discharge shots in the original data set. For example,
there are ten normal discharge shots in original data sets. The shot
numbers of ten normal discharge shots are 1-10. the training set is
\emph{shuffle}(1-6) (maybe one order is 1,4,6,5,2,3), validation set
is \emph{shuffle}(7-8), and testing set is \emph{shuffle}(9-10). Inner
a single normal shot discharge sequence will keep strict time order.

When all source data was obtained the z-scores \citep{Zill2011} will
be applied for standardization. And then all the preprocessed data
will be input to the deep learning model for training. In statistics,
the z-score is the number of standard deviations by which the value
of a raw score (i.e., an observed value or data point) is above or
below the mean value of what is being observed or measured. Raw scores
above the mean have positive standard scores, while those below the
mean have negative standard scores. z-score is calculated by $z=(x-\mu\text{)/\ensuremath{\sigma}}$
where $\mu$ is the mean of the population. $\sigma$ is the standard
deviation of the population.

The deep learning model uses an end-to-end training was executed on
8x Nvidia P100 GPUs with Keras \citep{chollet2015keras} and TensorFlow
\citep{abadi2016tensorflow} in the Centos7 system in local computing
cluster and remote computing cluster. The training of the deep learning
model starts with kernel initializer is glorot uniform initialization
\cite{Glorot2010}, the recurrent initializer is orthogonal \cite{Saxe2013},
bias initializer is zeros, and optimizer is Adadelta \cite{Zeiler2012}
for solving gradient explosion. The model trains about twelve days,
40 epochs. Then use callbacks and checkpoints to choose the best performing
model in $W_{mhd}$ and $n_{e}$ modeling best epoch is fifteen, while
$V_{loop}$ modeling best epoch is seven. In per epoch, all the data
in the training set will be put into the model for one time.

The training of our model is executed several times. Many of these
trials are considered as failed (e.g. divergence in training, poor
performance on the test set, etc) because of unsuitable hyper-parameters.
In the process of training our model, multiple sets of hyperparameter
are tried. Determine the best hyperparameter set by performance in
the validation set. Finally, the best hyper-parameters were found
by and shown in table \ref{tab:Hype}.

\begin{table}
\caption{Hyperparameters in this model \label{tab:Hype}}

\begin{tabular}{lll}
\hline 
Hyperparameter & Explanation & Best value\tabularnewline
\hline 
$\eta$ & Learning rate & $5\times10^{-3}$\tabularnewline
$\gamma$ & \selectlanguage{american}%
Adadelta\foreignlanguage{english}{ decay factor}\selectlanguage{english}%
 & 0.95\tabularnewline
Loss function & Loss function type & Mean squared error (MSE)\tabularnewline
Optimizer & Optimization scheme & Adadelta\tabularnewline
Dropout & Dropout probability & 0.1\tabularnewline
Epoch & Epoch & 15 and 7\tabularnewline
dt & Time step & 0.001s\tabularnewline
Batch\_size & Batch size & 1\tabularnewline
LSTM type & Type of LSTM & CuDNNLSTM\tabularnewline
LSTM size & Size of the hidden state of an LSTM unit. & 256\tabularnewline
$LSTM_{kernel}$ & Initializer for the kernel of LSTM weights matrix & Glorot uniform\tabularnewline
$LSTM_{recurrent}$ & Initializer for the recurrent kernel of LSTM weights matrix & Orthogonal\tabularnewline
Dense size & Size of the Dense layer. & 256\tabularnewline
$Dense_{kernel}$ & Initializer for the kernel of dense matrix & Glorot uniform\tabularnewline
$Dense_{bias}$ & Initializer for the bias vector & Zeros\tabularnewline
$n_{encoder}$ & Number of LSTMs stacked in encoder & 2\tabularnewline
$n_{decoder}$ & Number of LSTMs stacked in decoder & 2\tabularnewline
\hline 
\end{tabular}
\end{table}

\section{Results\label{sec:Validation}}

\begin{figure}
\includegraphics[width=0.5\textwidth]{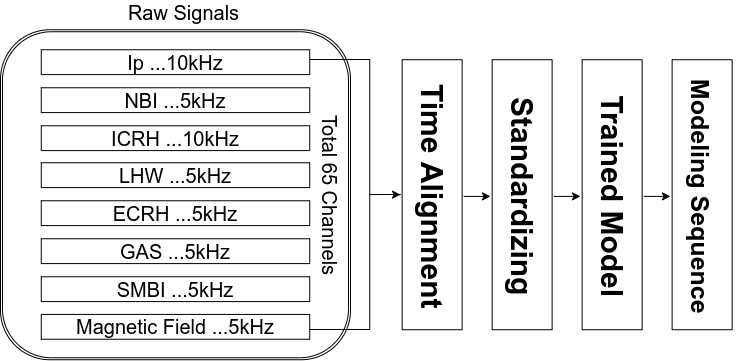}

\caption{Using the trained model.\label{fig:Usage}}
\end{figure}

After the deep learning model had been trained, the model can model
unseen data. As shown in figure \ref{fig:Usage}, at the beginning
of using the trained model, all data for 65 channel signals should
be obtained. In this step, It is necessary to keep and select the
same type of signal data in tokamak during training. The type of signal
data includes processed data and raw data. And then aligning all the
data on the time axis. The aligned data should be standardized by
the same parameters of the training set. All the standardized data
will be fed into the trained model and get the modeling sequence of
diagnostic signals, In the final step, the trained model should be
selected according to the diagnostic signal that you want to model.

In this section, the results of modeling will be analyzed in detail,
including representative modeling results and similarity distributions.
In this work, the similarity is a quantitative measurement of the
modeling results accuracy and is defined as follows:

\begin{equation}
S\left(\boldsymbol{x},\boldsymbol{y}\right)=\max\left(\frac{\Sigma(\boldsymbol{x}-\bar{\boldsymbol{x}})(\boldsymbol{y}-\bar{\boldsymbol{y}})}{\sqrt{\Sigma(\boldsymbol{x}-\bar{\boldsymbol{x}})^{2}\Sigma(\boldsymbol{y}-\bar{\boldsymbol{y}})^{2}}},0\right),
\end{equation}
where $\boldsymbol{x}$ is experimental data, $\boldsymbol{y}$ is
modeling result, $\bar{\boldsymbol{x}}$, $\bar{\boldsymbol{y}}$
are the means of the vector $\boldsymbol{x}$ and vector $\boldsymbol{y}$.

\begin{figure}
\includegraphics[width=0.3\paperheight]{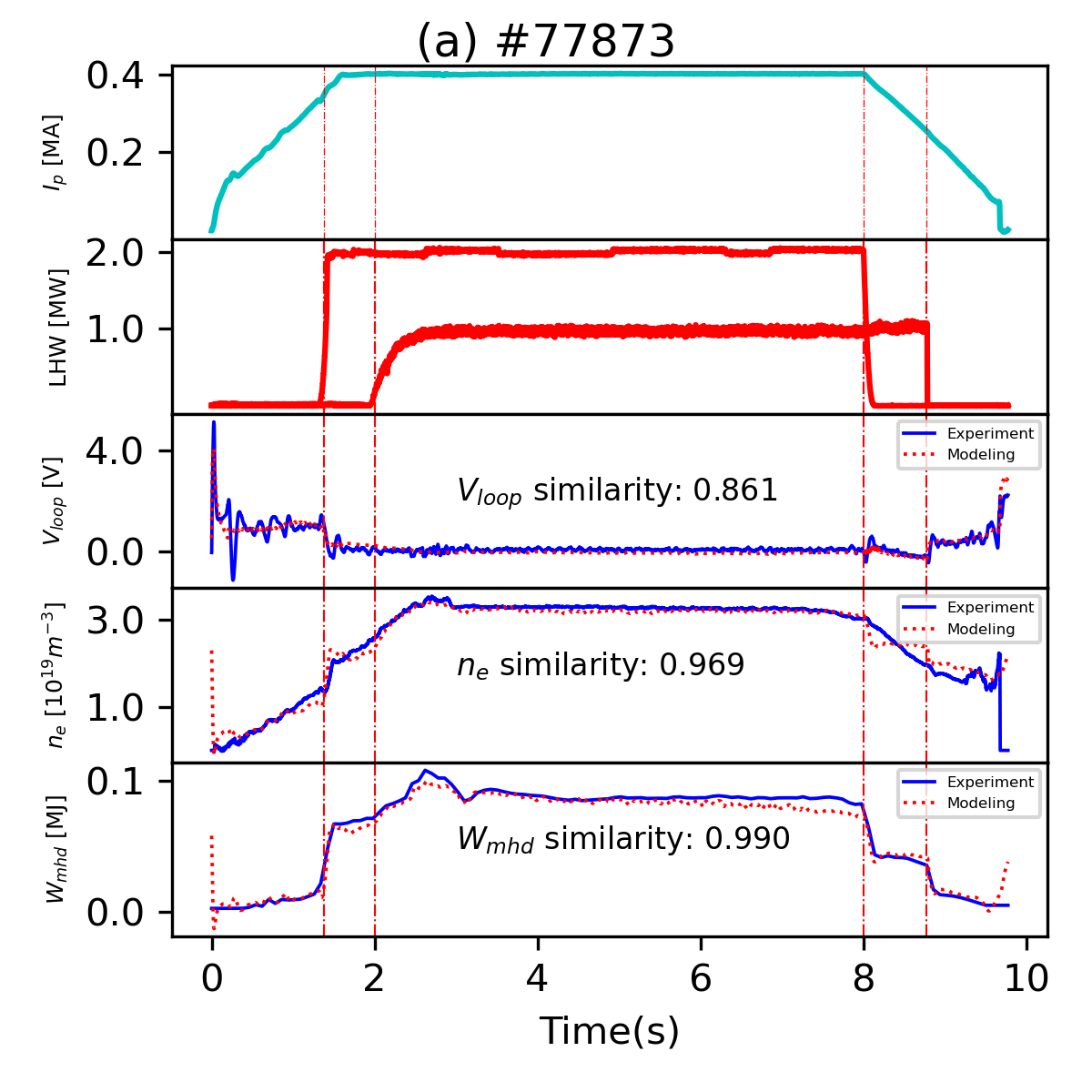}

\includegraphics[width=0.3\paperheight]{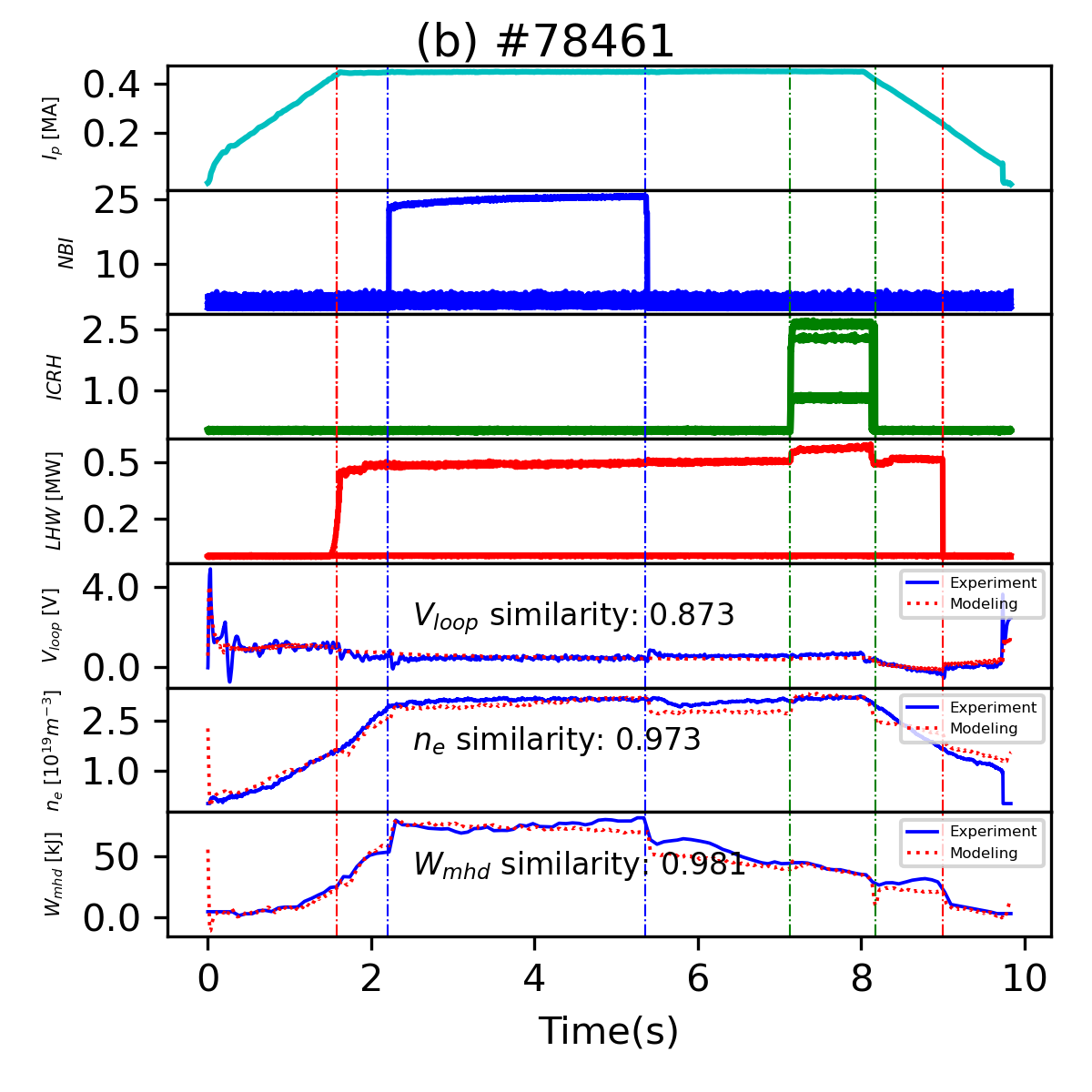}

\caption{Comparison of modeling result and EAST experiment data, shot \#77873
(a), and \#78461 (b). NBI and ICRH are the raw data so the physical
units are meaningless. \label{fig:Output}}
\end{figure}

Two typical EAST normal discharge shots shot \#77873 and \#78461,
are selected to check the accuracy of the model trained in this article.
Figure \ref{fig:Output}(a) shows the modeling result for shot \#77873,
which has two LHW injections during discharge. Figure \ref{fig:Output}(b)
shows the result for shot \#78461, which has NBI, LHW, ICRF injection.

Experimental data and modeling results are displayed together in figure
\ref{fig:Output}. The comparison shows that they are in good agreement
in most regions of discharge, from ramp-up to ramp-down. The slope
of the ramp-up and the amplitude of the flat-top are accurately reproduced
by the model. The vertical dash-dot lines indicate the rising and
falling edges of the external auxiliary system signal and the plasma
response, which show the time accuracy of the model.

Compared with experimental signals, the modeling results are more
sensitive to changes in external drives. For example, after the external
drive is turned off, the experimental signal $n_{e}$ continues to
decrease with a fixed slope, but the modeling results show a step-down.
However, it will also cause the deviation of the modeling result and
the experimental data when the external drive changes rapidly. How
to adjust the sensitivity of the model is still an open question.

\begin{figure}
\includegraphics[width=0.9\columnwidth]{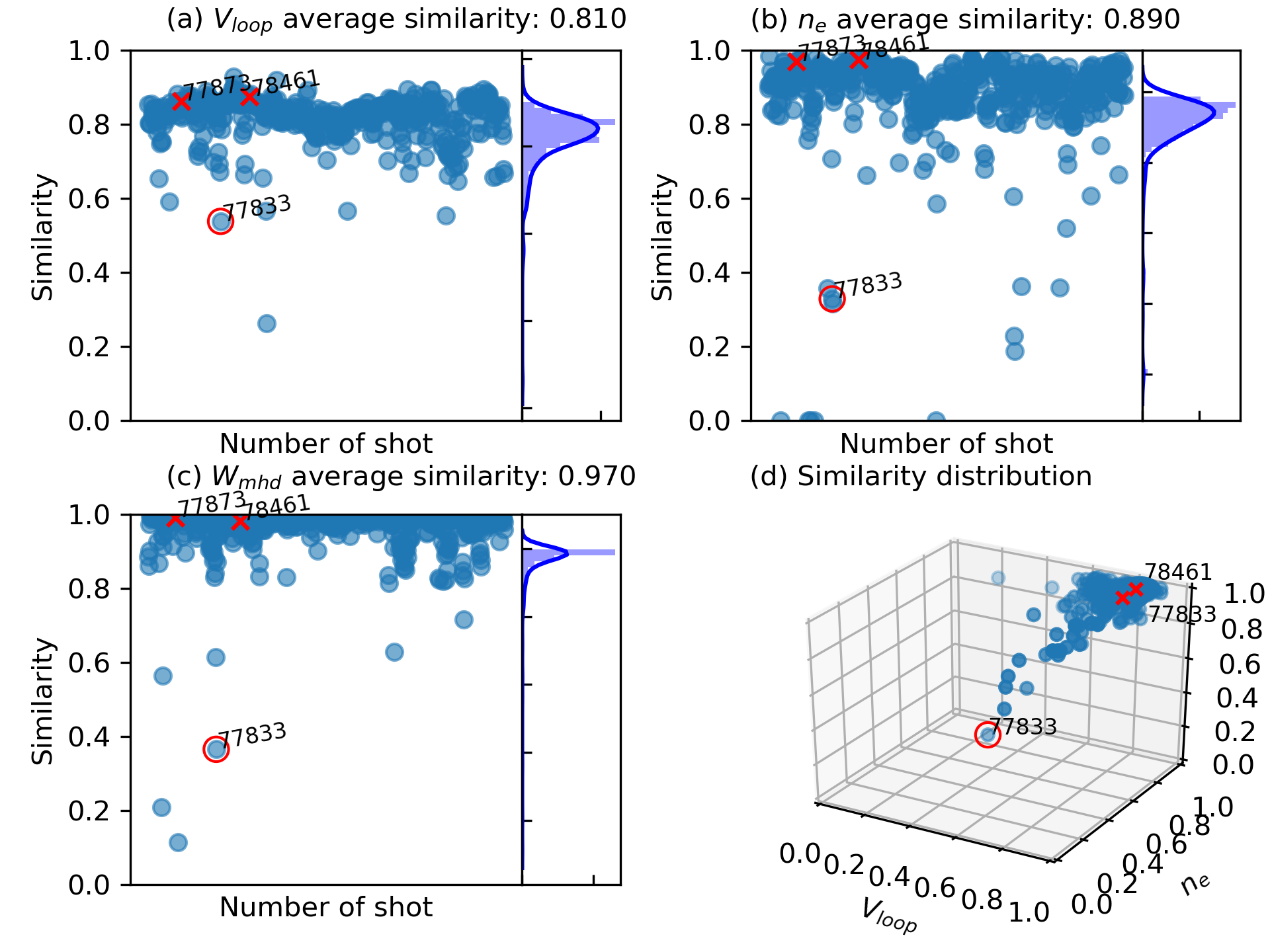}

\caption{The similarity distribution and average similarity in the test set.
show the similarity distributions of (a) $n_{e}$, (b) $V_{loop}$
, and (c) $W_{mhd}$, respectively. Figure (d) is a joint scatter
plot of three parameters. If the similarity is less than 0, it is
regarded as 0. \label{fig:Similarity-distribution}}
\end{figure}

A test data set with 695 shots were used to quantitatively evaluate
the reliability of the model. The statistical results of the similarity
between model results and the experimental data are shown in figure
\ref{fig:Similarity-distribution}.

$W_{mhd}$ is the best performance parameters, with the similarity
concentrated at more than 95\%. In other words, $W_{mhd}$ can be
considered to have been almost completely modeled under the normal
discharge condition. The almost similarity of $n_{e}$ is greater
than 85\%. $V_{loop}$ is the worst performing parameter, but many
of the errors are due to the plasma start-up pulse in the ramp-up
segment and the plasma shutdown pulse in the ramp-down segment. However,
$V_{loop}$ in the ramp-up and ramp-down sections is not the key factor
for the operation of the experiment.

\begin{figure}
\includegraphics[width=0.7\textwidth]{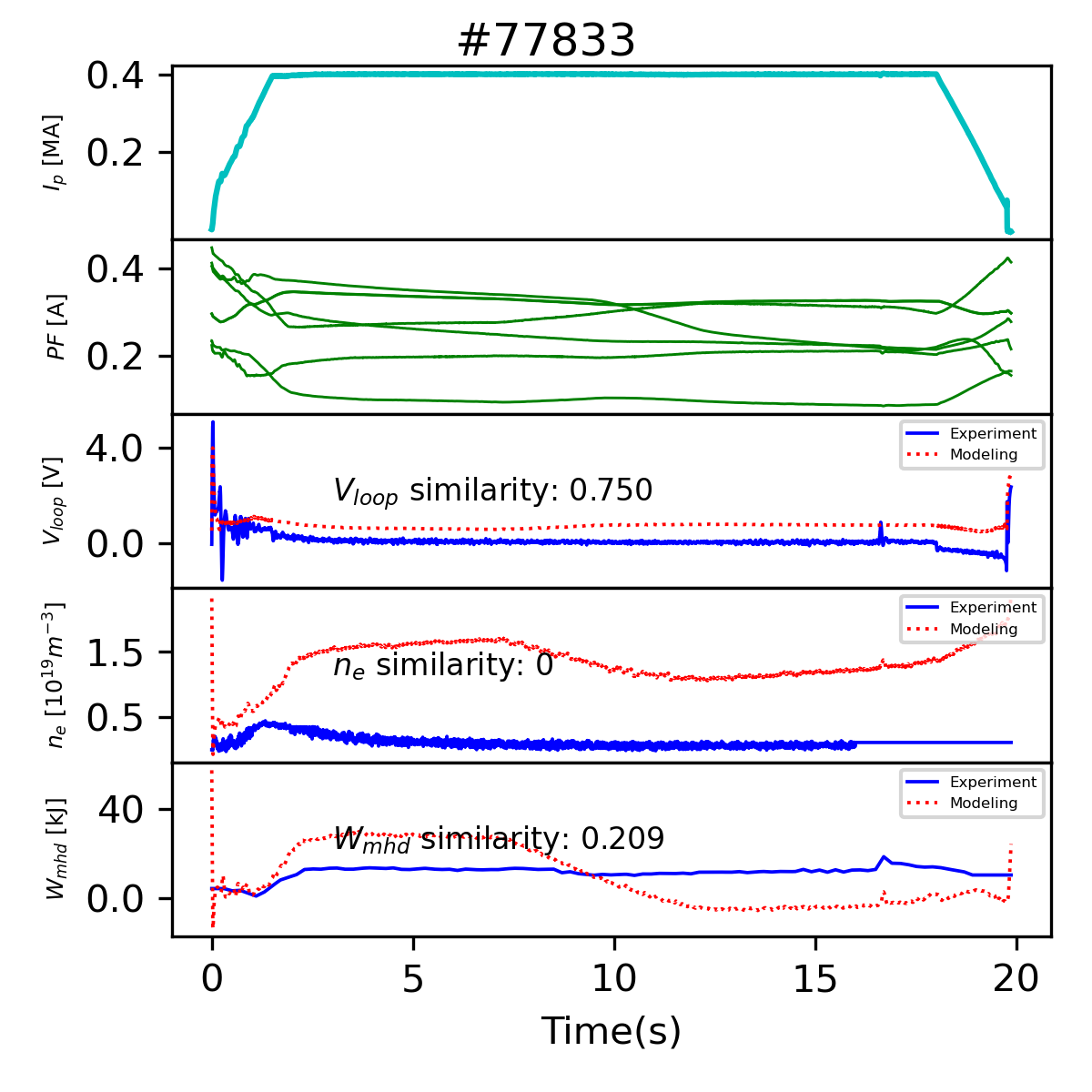}

\caption{shot \#77833 is a classical deviation caused by abnormal equipment
conditions. Shot \#77833 is used for cleaning device. \textcolor{red}{\label{fig:shot77833}}}
\end{figure}

The joint distribution of the three parameters is shown in figure
\ref{fig:Similarity-distribution}(d). Most shots are concentrated
in a limited range, which reflect the consistency of the model on
three target signals. It also shows that these shots belong to the
same tokamak operating mode. In other words, those points far away
from the center area indicate that the experiment is running in abnormal
mode. We checked all deviation shots in test set, and all deviation
are in abnormal conditions. For example,  shot \#77833 (as shown in
figure \ref{fig:shot77833}) is a classical deviation caused by abnormal
equipment conditions. This shot is used for cleaning device.

In terms of the similarity distribution of demonstrative parameters
and representative discharge modeling results, this machine learning
model application in tokamak discharge modeling is promising. $W_{mhd}$
can be regarded as almost completely reproduced under normal discharge
shot. $n_{e}$ can be successfully modeled in most areas under the
normal discharge condition. In normal discharge shot, the modeling
results of $V_{loop}$ at ramp-down and flat-top phases are in good
fitting with the experimental results.

\section{Conclusion \label{sec:Conclusion}}

In the present work, we showed the possibility of modeling the tokamak
discharge process using experimental data-driven methods. A machine
learning model based on the encoder-decoder was established and trained
with the EAST experimental data set. This model can use the control
signals (i.e. NBI, ICRH, etc.) to reproduce the normal discharge process
(i.e. electron density $n_{e}$, store energy $W_{mhd}$ and loop
voltage $V_{loop}$) without introducing physical models. Up to 95\%
similarity was achieved for $W_{mhd}$. Recent work of discharge modeling
has focused on physical-driven ``Integrated Modeling''. However,
this work shows promising results for the modeling of tokamak discharge
by using the data-driven methodology. This model can be easily extended
to more physical quantities, and then a more comprehensive tokamak
discharge model can be established.

Checking the physical goal of the experimental proposal is an important
and complicated problem. This work provides a reference for the realization
of physical goals under the normal discharge condition. Specifically,
the model mainly checks whether an experimental proposal can be achieved
under the normal discharge condition. Furthermore, if the experimental
result deviates greatly from the result of our model, there may be
two situations. one is that the experiment has some problems (as shown
in figure \ref{fig:shot77833}); another is that a new discharge mode
has appeared. The reason why a new discharge mode can be found is
that this model only models discharge mode that has appeared in EAST
normal discharges. Of note, if other discharge modes appear in the
data, this model will appear chaotic output as if it were abnormal
shots \citep{ISI:000355286600030}. In general, chaotic model output
means wrong input. For an experimental proposal, if the model gives
a chaotic output, the input should be carefully checked by an experienced
experimenter. If the input can be confirmed to be correct then it
is likely to be a new discharge mode. Our model is not capable of
fully accurate to recognize unsuccessful sets of inputs.

Compared with the physical-driven method, the data-driven method can
build models more efficiently. We also realize that there are many
challenges before the practical application of this method. For example,
the impact of model sensitivity on modeling results has been recognized.
How to adjust the sensitivity of the model is still an open question.
Cross-device modeling is more important for devices under design and
construction such as ITER and CFETR. Introducing device configuration
parameters and performing transfer learning is a feasible solution
to this problem. Our next step is to model the time evolution of the
one-dimensional profile and the two-dimensional magnetic surface.

\ack{}{}

The author would like to thank all the members of EAST Team, especially
Feng Wang, for providing such a large quantity of past experimental
data. The author Chenguang Wan sincerely thanks Yong Guo, Dalong Chen
for explanation of the experimental data, Cristina Rea, and Professor
Robert Granetz for technical discussion.

This work was supported by the National MCF Energy R\&D Program under
Contract No.2018YFE0304100 and the Comprehensive Research Facility
for Fusion Technology Program of China under Contract No. 2018-000052-73-01-001228.

\bibliographystyle{unsrt}
\bibliography{Reference_database}

\end{document}